\documentclass[showpacs,aps,prl,twocolumn]{revtex4}
\usepackage{graphicx,amsmath}
\usepackage[dvips]{color}
\newcommand{\beq}{\begin{equation}}
\newcommand{\eeq}{\end{equation}}
\newcommand{\bea}{\begin{eqnarray*}}
\newcommand{\eea}{\end{eqnarray*}}
\newcommand{\bnea}{\begin{eqnarray}}
\newcommand{\enea}{\end{eqnarray}}
\newcommand{\p}{\partial}

\newcommand{\ba}{\begin{array}{cc}}
\newcommand{\ea}{\end{array}}
\newcommand{\bm}{\left[ \begin{array}{cc}}
\newcommand{\fm}{\end{array} \right]}

\begin{document}

\title{Convective instability induced by nonlocality in nonlinear diffusive
systems }
\author{Francesco Papoff and Roberta Zambrini}
\affiliation{Department of Physics, University
of Strathclyde,  107 Rottenrow,  Glasgow G4 0NG, UK.}
\date{\today}

\begin{abstract}

We consider a large class of nonlinear diffusive systems with nonlocal
coupling. By using a non-perturbative analytical approach we are able to
determine the convective and absolute instabilities  of all the uniform states
of  these systems.
We find a huge window of convective instability that  should
provide a great opportunity to study experimentally and theoretically noise
sustained  patterns. 

\end{abstract}

%42.65 Sf  Dynamics of nonlinear optical systems; optical instabilities, optical chaos and complexity, and optical spatio-temporal dynamics
%05.40.ca  Noise
%89.75.Kd  Patterns
%05.45.-a  Nonlinear Dynamics and nonlin. dyn. systems

\maketitle

There has  been a considerable interest in convective instabilities and noise
induced patterns recently~\cite{santagiustina97-taki00}. A convective
instability happens when a state of a nonlinear system becomes unstable  and  a
localised perturbation  grows travelling in the system, but  eventually decays
at any point in the laboratory frame.  In this regime a state  different from
the original one cannot be established   unless it is sustained by noise. Small
regions of convective instabilities have been predicted and observed  in
hydrodynamics~\cite{ahlers83-gondret99}, plasma physics \cite{briggs} and 
optics~\cite{Louvergeaux04}. These are due to spatial drift terms modelled by
gradients in the direct space or by the first few terms of the  Taylor
expansion of the dispersion relation in the Fourier 
space~\cite{Louvergeaux04}.

In this letter, instead, we study  nonlinear systems with nonlocal coupling.
Nonlocality is fundamental in modelling material response in presence of 
transport effects~\cite{nonlocalrefs}. Even for local material response,
nonlocality  is  unavoidable whenever travelling waves emerging from a medium
are reflected back to it non-collinearly, therefore coupling any spatial point
$x$ with the shifted point $x+\Delta x$. Indeed this kind of nonlocal coupling
is induced by any small misalignment in all feedback optical
systems (see experiments in Refs.~\cite{ramazza98,Louvergeaux04,thorsten99}).
We consider the most general case in which the spatial shift cannot be
approximated by a gradient (drift) term in a very  broad class of nonlinear
diffusion equations. Equations of this type arise in nonlinear systems with
diffraction-free optical feedback \cite{ramazza98} and are  an interesting
generalisation of more standard nonlinear diffusion equations.
%  common in many other
%fields~\cite{Murray-Ortoleva}.
We show that the stability of all the uniform states of this class of systems 
is determined by a single dispersion relation with two parameters.
By using a non-perturbative analytical approach we are able to
analyse this dispersion relation. We find a huge 
convective instability window where noise sustained patterns 
and amplification of perturbations can be observed.

We consider equations of the type 
\bnea
( \partial_{t} -\partial_{x}^2) \phi(x,t)  = f_1(\phi(x,t);\mu) + 
%T_{\Delta x} \circ 
f_2(\phi(x+\Delta x,t);\mu) \label{model}
\enea
where $\phi$ is a real variable, $t$ is in units of the diffusion time, 
$x$ and the spatial shift $\Delta x$ are in unit of the diffusion length, 
$\mu$ is a control parameter independent on $x$.  
$f_1$, $f_2$ are real functions that can be derived with respect to 
$\phi$. The uniform states $\phi_{m}$ of Eq.(\ref{model})
are the solutions of $f_1+ f_2=0$ and their domains of existence   
depend upon $\mu$ but not upon $\Delta x$ in the limit of 
infinitely extended systems. 
The dispersion relation for perturbations 
$ \exp{(\omega t +i k_I x)}$ of a uniform state of Eq.~(\ref{model})
 is 
\beq
\omega=-k_I^2 +\p_{\phi}f_1(\phi_{m};{\mu})+ 
\p_{\phi}f_2(\phi_{m};{\mu}) e^{ik_I\Delta x} \label{disp_Rel}.
\eeq
As a consequence of the term $e^{ik_I\Delta x}$, which is present in
all systems with  shift,
there are bands of $k_I$ for which the real part of the dispersion relation
$\omega_R$ can be positive. For  $\p_{\phi}f_1(\phi_{m};{\mu})<0$, as in the
experiments in Ref~\cite{ramazza98}, these bands are within the regions
where $\p_{\phi}f_2(\phi_{m};{\mu}) \cos{k_I\Delta x}>0$.
As a result the homogeneous solution $\phi_{m}$ is unstable and
plane wave perturbations are amplified. 
As the imaginary part of $\omega$ (phase velocity) is in general non null, 
these waves move across the system.
A peculiar effect of the non-locality is that for
spatially localised perturbations the sign of
the group velocity and of the phase velocity of the most unstable wavenumber
are always $opposite$.

\begin{figure}
  \centering
  \includegraphics[width=8cm]{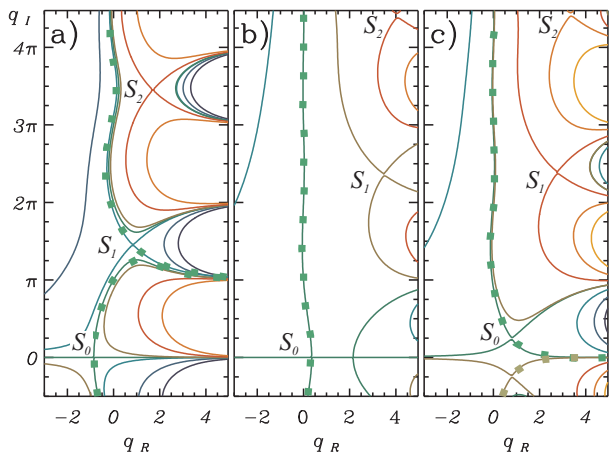}
  \caption{Equiphase paths in the complex plane, $w_I (q_R, q_I)=$ constant,
with $w= \omega \Delta x^2$ and   $q=k \Delta x$, for 3 values of the control
parameter $ \p_{\phi}f_2(\phi_{m};{\mu})\Delta x^2 = 4$ (a), $-0.5$ (b), $-1$
(c).  The saddles points $s_i$ are at the intersections between equiphase
paths. The symbols show the steepest descent paths closing the integration
contour.}
      \label{fig:saddles}
\end{figure}

When the group velocity is non vanishing, 
we have to consider whether localised perturbations 
produce absolute or convective instability, that is, to find whether
the Green function,$ \int^{+\infty}_{-\infty}e^{ik_Ix+\omega(k_I)t}dk_I$, of the 
linearised equation diverges or vanishes for 
$t\rightarrow +\infty$~\cite{Infeld_Rowlands}.
To evaluate the integral, it is convenient to extend analytically  
$\omega$ in the complex plane $k=k_R+ik_I$ and apply the saddle-point method. 
 An even number (at least
four) of paths with $\omega_I$ constant start from each saddle point 
with $d\omega/dk=0$, as shown in Fig.~\ref{fig:saddles}.
On half of these equiphase paths, the steepest
descent, $\omega_R$ decreases fastest; on the remaining half, the steepest ascents,
$w_R$ increases fastest. {\it If} one can form a closed integration contour with
the imaginary axis and steepest descents, then the asymptotic value of the 
integral is given by the values of $\omega$ {\it at the saddle points on
 the integration contour}~\cite{bender99}. 
This method has been extensively used to find the threshold between
convective and absolute instabilities in systems with drift, where the
dispersion has in general few saddle points.    However for nonlocal 
systems the exponential term in the dispersion originates
always  a countable infinity of saddle points.
Moreover, the saddles and their steepest  descents move and can suddenly
collide and disappear as the control parameters change.
It is therefore essential to study very carefully how the global 
geometrical organisation of the saddles and of the 
equiphase paths changes with the control parameters in order to close 
correctly the integration contour. 

In order to simplify the calculations, we define the parameters 
$\xi=\p_{\phi}f_1(\phi_{m};{\mu}) $, $\chi=\p_{\phi}f_2(\phi_{m};{\mu}) $. The
analytic extension of the dispersion relation is  $w=\xi \Delta x^2  + q^2 +
\chi \Delta x^2  e^{q}$,  with  $w= \omega \Delta x^2$ and $q=k \Delta x$, and
we consider only the semi-plane $q_I \ge 0$ because $w(q)=w^{*}(q^{*})$. Note
that the dispersion relation and the stability depend only upon  the  effective
parameters $\xi \Delta x^2$ and $\chi'\equiv \chi\Delta x^2 $. For all values
of $\chi'$ there is at least a saddle and at most two saddles in the interval
$q_I \in [0,\pi)$. We call the leftmost of these saddles $s_0$ (see
Fig.~\ref{fig:saddles}b).   For all $n>0$, there are also saddles $s_n$ in the
$n$-th interval  $q_I \in (2n\pi,(2n+1)\pi)$ for $\chi'<0$
(Fig.~\ref{fig:saddles}b-c)  or in the $n$-th interval  $q_I \in
((2n-1)\pi,2n\pi)$ for $\chi'>0$ (Fig.~\ref{fig:saddles}a). In order to  close
the integration contour entirely with steepest descents, we need a steepest
descent path that ends on or is asymptotic to the real axis  ($q_R$) and
another asymptotic to the imaginary axis ($q_I$). These steepest descents must 
either come from the same saddle or be connected by other steepest  descents.

From the equation of the equiphase paths that originate at the saddle $s$, 
$2q_Iq_R+\chi'e^{q_R}\sin{q_I}=w_I(s)$, we can show that the possible
asymptotes of the steepest descents are the  imaginary axis for $q_I
\rightarrow +\infty$ and the lines $q_I=(2n+1)\pi$ for $\chi'>0 $  and $n\geq
0$ (or the lines $q_I = 2n \pi$ for $\chi'<0$) and  $q_R \rightarrow +\infty$. 
We determine the geometrical organisation of the steepest descents by using
this information, together  with the exact determination of the interception of
the  steepest descents with the lines  $q_I = n\pi$ and with the imaginary axis
and the fact that, in this problem,   two steepest descents with different
phases can  intersect only at infinity~\cite{Arfken}.  We find analytically
that the saddle with the smallest phase $w_I$  ($s_1$ in 
Fig.~\ref{fig:saddles}a and $s_0$ in Fig.~\ref{fig:saddles}b and  c) has
always  the upper steepest descent to the left of all the other steepest
descents and  asymptotic to the imaginary axis.  Indeed this saddle is always
necessary to close the integration contour. If $s_0$ happens to be the saddle
with the smallest phase, then the steepest descents of $s_0$ close the
integration contour  (Fig.~\ref{fig:saddles}b-c) because $s_0$ either lies on
the real axis or has another steepest descent asymptotic to the real axis. If
instead  there is another  saddle $s_{n_1}$ with $w_I(s_{n_1}) \le w_I(s_0)$
(for instance the saddle  $s_1$ in Fig.~\ref{fig:saddles}a), than the upper
steepest descent from $s_0$ remains below  $s_{n_1}$ and is connected to the
steepest descent from  $s_{n_1}$ (in Fig.~\ref{fig:saddles}a these two steepest
descents are asymptotically connected at $q_I=\pi$).  In this case we use a
steepest descent  from $s_{n_1}$ to reach values of $q_I$ above  $s_{n_1}$
itself and a  steepest descent from $s_0$ to reach the real axis. If $s_{n_1}$
is the saddle with the smallest phase, then the steepest descents of $s_0$ and
$s_{n_1}$ close the integration  contour (symbol lines in
Fig.~\ref{fig:saddles}a),  otherwise we will have to include also the first
saddle, $s_{n_2}$, with $n_2>n_1$ and $w_I(s_{n_2}) \le w_I(s_{n_1})$. In order
to find the saddles to be used  to close the integration contour, we repeat
this process including all the saddles whose phase is the minimum of the phases
of the saddles below them. For each finite $\chi'$,  this procedure gives a
finite set  of saddles $P_{\chi'}$ 
\footnote{Given that $w_I(s_{0}) \le 0$ and that, if 
$w_I(s_{n_{\chi'}}) > 0$, we have $w_I(s_{n}) > 0$ for all 
$n>n_{\chi'}$, we can define this set as 
$P_{\chi'}= \{s_{n}|~0 \le n \le n_{\chi'} , w_I(s_{n})= 
\min_{0 \le j \le n} \{w_I(s_j)\}\}$.
}.
Examples of different integration contour are shown in Fig.(\ref{fig:saddles}) as
symbols lines.

\begin{figure}
  \centering
  \includegraphics[width=7.5cm]{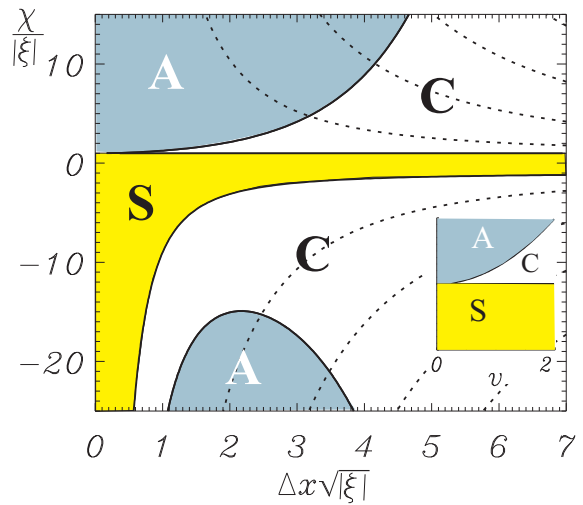}
  \caption{Instabilities diagram: Stable (S), convectively unstable (C)
  and absolutely unstable (A) regions for $\xi < 0$.
  The dashed lines indicate where different bands become unstable. 
  The insert shows the well-known
  diagram \cite{deissler} obtained when 
  instead of a non-local coupling a  walk-off term ($v$) is considered.
  In this case there are no instabilities at all in the lower diagram region.}
      \label{fig:thresholds}
\end{figure}

From Eq.~(\ref{disp_Rel}) we identify the region  $w_R(0,q_I)<0$ in which the 
homogeneous solutions are stable.
This is the region S in Fig.~\ref{fig:thresholds} and its contour is the 
convective threshold. 
If $w_R(0,q_I)>0$ for some $q_I$ and if at least one saddle 
$s_n \in P_{\chi'}$ has $w_R(s_n)>0$,
 the instability is absolute. 
However, we  need to find only a small subset of saddles in  $P_{\chi'}$
to determine the stability because 
$w_R(s_n) > 0$ {\it only if there is instability with $w_R(0,q_I)>0$
for $q_I$ in the $n$-th interval} 
\footnote{This follows from the fact that, for $s_n \in P_{\chi'}$,
$w_R(s_n) > 0$ only if  ${s_n}_R < 0$ and that, 
for ${s_n}_R < 0$, there is a
path with $w_R$ constant that connects the saddle with a point on the 
imaginary axis within the $n$-th interval.}. 
Therefore, for each $\mu$, we can determine the nature of the 
instability by finding the values of $w$ at the saddles in $P_{\chi'}$ that are 
in bands  with instability threshold ${\chi'}^c_n$  below the threshold ${\chi'}^a_0$
where $w_R(s_0)=0$. The absolute threshold 
$\chi'^a$ is then the value of $\chi'$ where
\beq
\max\{w_R(s_n)|s_n \in P_{\chi'}, {\chi'}^c_n \le {\chi'}^a_0\}=0. \label{abs_th}
\eeq

The importance of the determination of $ P_{\chi'} $ and of Eq.~(\ref{abs_th})
is twofold. On one hand they guarantee that we can apply 
the  method of steepest descents by  properly closing the integration contour; 
on the other hand, they allow us to find the  
absolute threshold simply by inspection of $w$ at a finite number of saddles. 
This is remarkable in view of the  infinite number of saddles 
produced by the shift.
Moreover,  $P_{\chi'}$ is the same for all 
systems with the same $\chi'$ in the class considered because the global 
geometrical organisation of the saddles and of their steepest descents 
 does not depend on $\xi$.

Applying  this technique we obtain the instabilities diagram in 
Fig.~\ref{fig:thresholds} valid for any $\xi \le 0$ 
(including the typical case of a linear damping term $\xi =-1$). 
For $\chi < 0$ the instability bands  have $q_I\neq 0$ and the lowest convective
threshold is very far from the absolute threshold. For $\chi>0$, the 
instability with respect to perturbations with $q_I=0$ is absolute for $\xi=0$ 
and convective for $\xi \ne 0$,  with the convective instability 
windows increasing as $\xi$ decreases. 
Comparing the instabilities diagram for diffusive problems 
with finite shift and with with walk-off \cite{deissler} 
(Fig.~\ref{fig:thresholds}) we note that
the whole $modulation$ instability region for negative values of
$\chi$ is a specific 
effect of a finite shift, as was already recognised by \cite{ramazza98}.  
 What our analysis reveals by direct calculation of the absolute threshold
 is that in this parameter region the system is mainly $convectively$ 
 unstable and shows noise sustained modulated
 patterns. Only for very negative values of $\chi/|\xi|$ the absolute threshold
 is crossed.

\begin{figure}
  \centering
  \includegraphics[width=4cm]{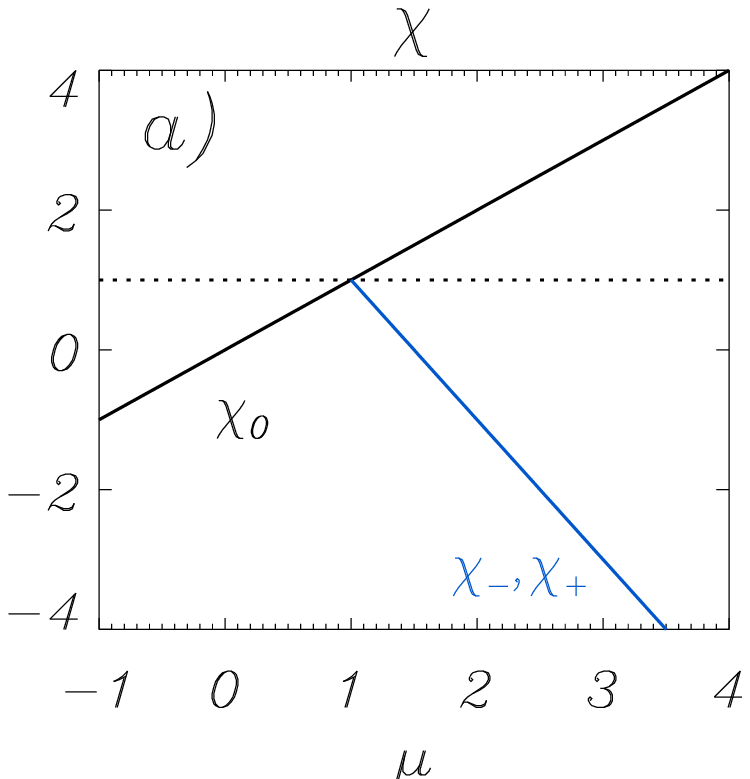}
  \includegraphics[width=4cm]{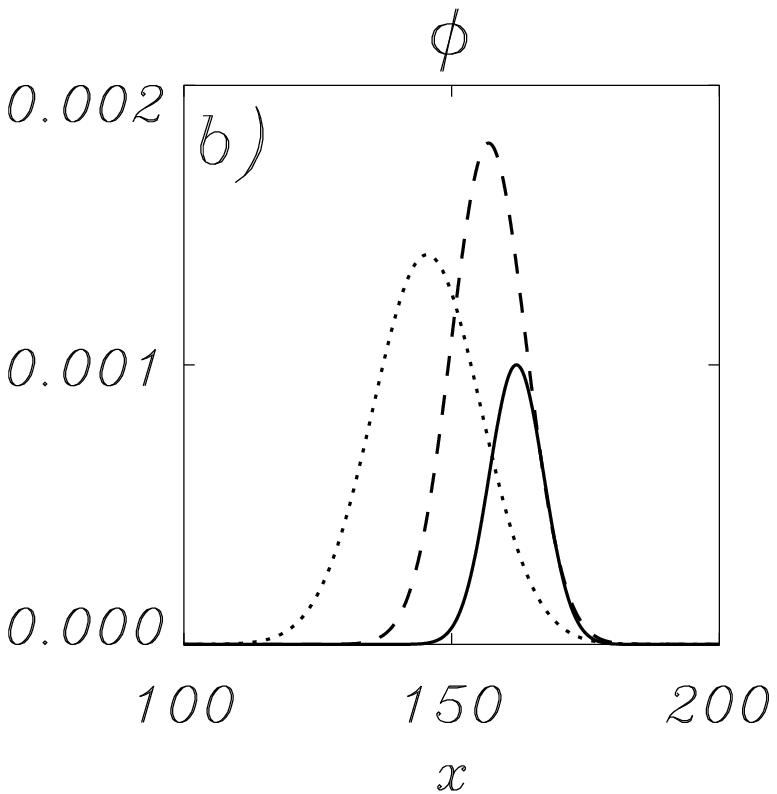}
  \caption{ a) $\chi_0(\mu)$ and $\chi_\pm(\mu)$ for the  NGL equation.
   b) Gaussian perturbation (continuous line) of the vanishing state 
   $\phi_0$ and evolved state for $\mu=1.03$ (dotted line)
     and for $\mu=1.09$  (dashed line).  Shift $\Delta x=0.48$. 
     The instability for $\mu=1.03$ is convective
     (local decay, even if the maximum of the perturbation grows),
     while for $\mu=1.09$ it is absolute
     (the perturbation grows also locally).
     After a longer transient the perturbation ends up colliding with 
     the uniform stable states
     $\phi_+$ or $\phi_-$.
     Simulations with space discretization $0.12$, $4096$  grid points and
     time step $0.001$.
     }     
\label{fig_b_mu}
\end{figure}
To show the general applicability of the stability analysis presented in this
letter we consider in the following two examples: the Ginzburg-Landau and the
saturable nonlinear equations, both with nonlocal nonlinear terms.
A nonlocal Ginzburg-Landau (NGL) equation is obtained by Eq.~(\ref{model}) when
\begin{equation}
\label{eq1}
f_1= - \phi, ~~~~f_2= \mu \phi-\phi^3.
\end{equation}
The three uniform states $\phi_0=0$ (for any $\mu$) and  $
\phi_\pm=\pm\sqrt{\mu-1}$  (for $\mu>1$) are associated to the  
parameters $\xi_0=\xi_\pm=-1$ and $\chi_0=\mu$, $\chi_\pm=3-2\mu$ 
(Fig.~\ref{fig_b_mu}a).  Given the relation between $\chi_{0,\pm}$ and  $\mu$,
Fig.~\ref{fig:thresholds} provides the linear stability diagram of each state
of the NGL equation. In particular, $\chi_{\pm}$ is always less than 1 so that
only the lower part of the diagram in Fig.~\ref{fig:thresholds} describes  $
\phi_\pm$ instabilities, while for the state  $\phi_0=0$
all the instabilities shown in the diagram arise by varying $\mu$ and the shift
$\Delta x$.  Without shift the state  $\phi_0$ becomes unstable for 
$\mu>1$ where the system evolves to the homogeneous states 
$\phi_+$  and $\phi_-$(connected by fronts). With a
finite shift for $\mu>1$ a region of convective instability arises before the
absolute threshold.  Simulations of the NGL equation 
with small values of $\mu$ show instabilities of
the states $ \phi_{0,\pm}$ in good agreement with our theoretical
analysis. Fig.~\ref{fig_b_mu}b is an illustrative example 
of the typical evolution
of an initial perturbation of the unstable state  $\phi_0$, below and
above the predicted  absolute threshold.

\begin{figure}
  \centering
  \includegraphics[width=3.6cm]{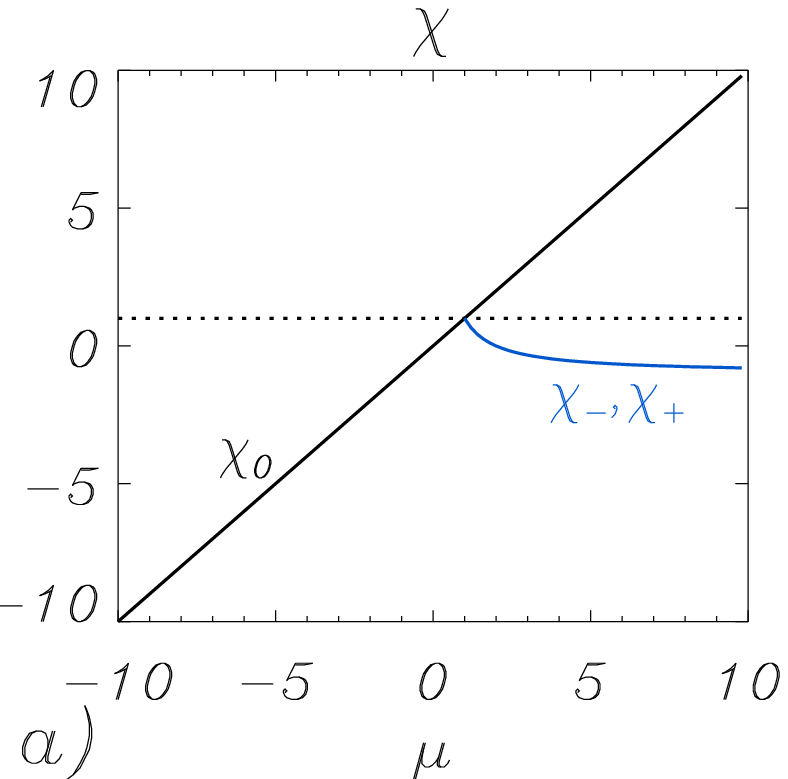}
  \includegraphics[width=4.5cm]{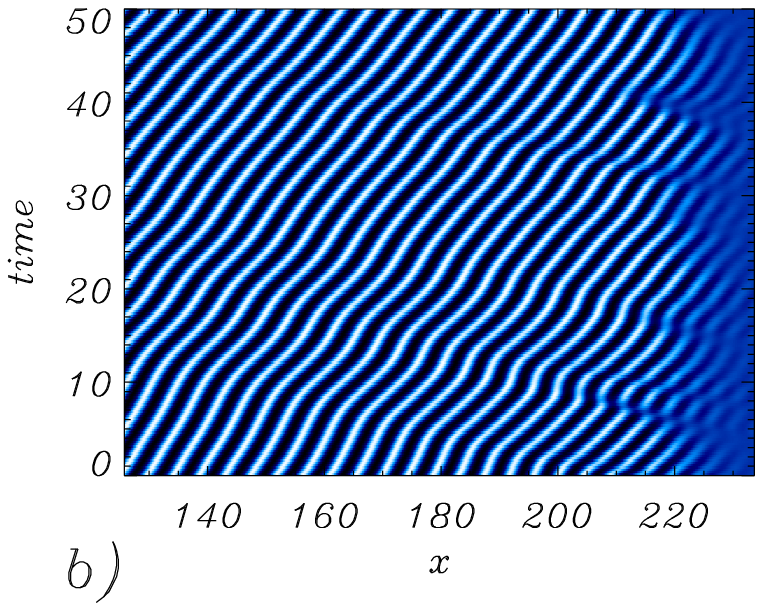}
  \caption{ a) $\chi_0(\mu)$ and $\chi_\pm(\mu)$ for the  AS equation.
   b) Noise sustained pattern
   for $\chi_0/|\xi|=\mu=-5$ and $\Delta x=1.92$ in the NAS with
   an additive Gaussian white noise of amplitude $0.01$.
    Simulations with Dirichlet boundary
   conditions and same numerical discretization of Fig.~\ref{fig_b_mu}.}     
\label{fig_b_mu2}
\end{figure}
We now consider the nonlocal saturable nonlinear (NSN) equation obtained by 
Eq.~(\ref{model}) when
\begin{equation}
\label{eq2}
f_1= - \phi, ~~~~f_2= \mu \frac{\phi}{1+\phi^2}.
\end{equation}
We find the same three uniform states $\phi_0=0$, 
$ \phi_\pm=\pm\sqrt{\mu-1}$ as before, but now
$\chi_0=\mu$, $\chi_\pm=\frac{2-\mu}{\mu}$ 
(Fig.~\ref{fig_b_mu2}a). 
Again, knowing $\chi_{0,\pm}(\mu)$ and $\xi=-1$, Fig.~\ref{fig:thresholds}
gives the thresholds of the NSN in terms of the specific 
parameters $\mu$ and  $\Delta x$.
We have seen before that 
for negative values of $\chi$ (here $\chi_{0}=\mu<0$) new 
modulation instabilities  are predicted
(both convective and absolute),
existing $only$ for non-vanishing shifts and not observed
in systems with drift. In Fig.~\ref{fig_b_mu2}b we show an example of a noise
sustained  stripe pattern, as predicted by our
theoretical analysis.

\begin{figure}
  \centering
  \includegraphics[width=8cm]{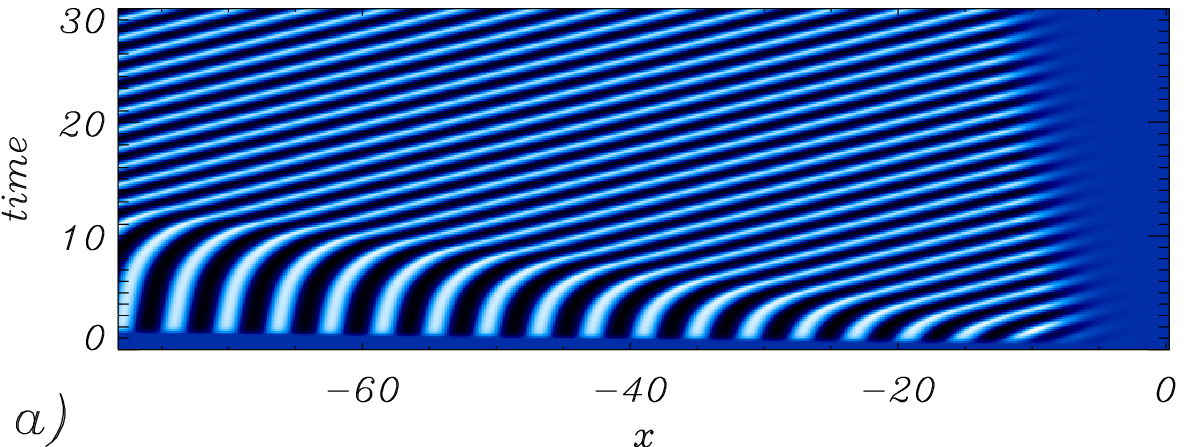}
  \includegraphics[width=8cm]{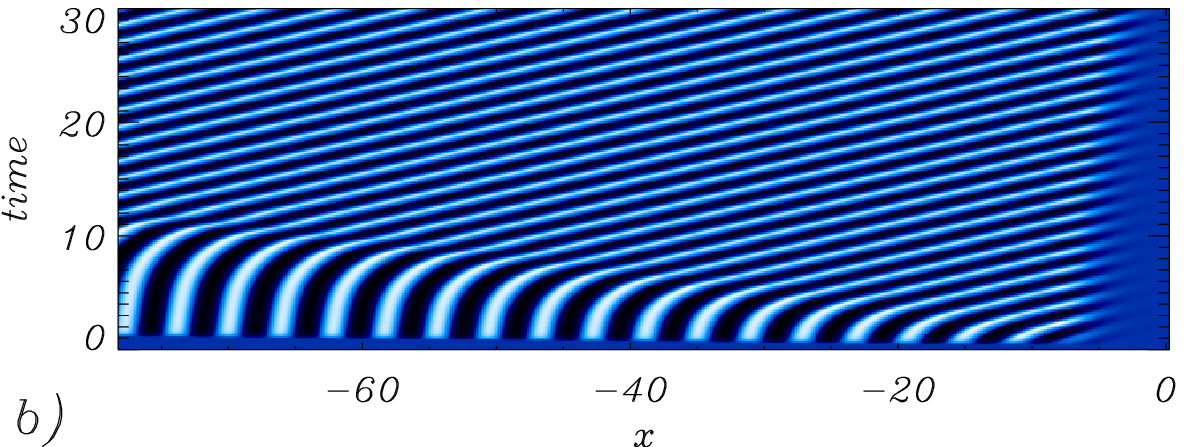}
  \caption{Temporal evolution of an initial Gaussian in $x=0$
  perturbing the vanishing state $\phi_0$.
  a) For $\mu=-14$  the
  perturbation evolves toward a modulated wave-packet that eventually leaves 
  the system (convective instability).   b) For $\mu=-18$ the right front moves to
  the positive $x$ space region, and finally the modulated solution occupies the
  whole system (absolute instability).
  Simulation of the deterministic NSN equation with same 
  parameters as in Fig.~\ref{fig_b_mu2}b.}     
\label{pulse}
\end{figure}

The NSN equation has the advantage of not developing divergences
 even for large values of the
control parameter due to the saturable nonlinearity. 
Therefore it was possible to check the instability diagram even for large
negative values of $\chi_0$, decreasing $\mu$  from $-14$ to $-18$ 
(Fig.~\ref{pulse}). Due to the modulational character of the instability an
initial Gaussian perturbation does not evolves in the simple way shown in
Fig.~\ref{fig_b_mu}b. Nevertheless from the temporal evolution of the fronts it
is still possible to distinguish without ambiguity  between convective 
(Fig.~\ref{pulse}a) and absolute (Fig.~\ref{pulse}b) instabilities. It is also
interesting to note that the wave-packet shows a clear asymmetry between the
left and the right edges,  with the wavenumbers with lower phase
velocity in the leading edge and those with higher phase velocity in the
trailing edge. This effect results from the
symmetry  breaking caused by the shift and the opposite sign of the group
velocity and the phase velocities of most wavenumbers. 
%Old:--
%where different wave-vectors with different phase
%velocities are selected. This effect results from shift breaking reflection
%symmetry  and  it is observed in presence of fronts (for instance in
%Fig.~\ref{pulse} the front connecting the vanishing and the modulated 
%solutions).

%
%

%For the first time in our knowledge, we have shown that 
%nonlocal systems  develop conv inst.

In conclusion we have identified nonlocality  as a new mechanism leading to
convective instabilities in a large class of systems. We have shown how to
determine the absolute threshold by finding the values  of $w$ at a finite
number of saddles which are selected out of an  infinite number by the
procedure described here. Our non perturbative approach allows us to analyse
situations in which there are several bands of unstable wavenumbers and the
absolute instability is very far from the lowest convective threshold. 
Theoretical  predictions have been confirmed by numerical simulation of the
dynamics of two prototype models. 
%This analysis could be generalised to system
%with temporal delay.  
We expect that experiments in nonlocal systems  will
exhibit a dynamics dominated by noise and boundary  effects in a very large
region of control parameters.  This opens the possibility of new and exciting
research in the fundamental properties of extended nonlinear systems.

We acknowledge discussions with  M. San Miguel, D. Walgraef and E. Yao.
RZ is financially supported by the UK Engineering and 
Physical Sciences Research Council (GR/S03898/01).

%%%%%%%%%%%%%%%%%%%%%%%%%%%%%%% end of text %%%%%%%%%%%%%%%%%%%%%%%%%%%%%%%

\end{document}